\documentclass[3p, 12pt, sort&compress]{elsarticle}

\usepackage{upgreek}
\usepackage{soul}

\pdfoutput = 1

\usepackage[font={it}]{caption}

\usepackage{dsfont, bm, 
  amssymb, amsmath,
  graphicx,
  hyperref,
  array,
  subfig,
  upgreek,
}

\journal{Nuclear Physics B}

\newcommand{\mat}[1]{\bm{#1}}

\newcommand{\D}[2]{\mathrm{d}^{#1}{#2}}

\begin{document}

\begin{frontmatter}

\title{{\bf {\LARGE Corrigendum to \\``Flavour Covariant Transport Equations:\\
an Application to Resonant Leptogenesis''}}\medskip}

\author[a]{\large P.~S.~Bhupal Dev}

\author[b]{\large Peter Millington}

\author[a]{\large Apostolos Pilaftsis}

\author[a]{\large Daniele Teresi}

\address[a]{Consortium for Fundamental Physics,
  School of Physics and Astronomy, \\ 
  University of Manchester, Manchester M13 9PL, United Kingdom.}

\address[b]{Physik Department T70, James-Franck-Stra\ss e,\\
Technische Universit\"{a}t M\"{u}nchen, 85748 Garching, Germany.}
\begin{abstract}
We amend the incorrect discussion in Nucl.~Phys.~B \textbf{886} (2014) 569
\cite{Dev:2014laa} concerning the numerical examples considered there. In particular,
we discuss the viability of minimal radiative models of Resonant
Leptogenesis and prove that no asymmetry can be generated at
$\mathcal{O}(h^4)$ in these scenarios. We present a minimal
modification of the model considered in \cite{Dev:2014laa}, where electroweak-scale
right-handed Majorana neutrinos can easily accommodate both successful
leptogenesis and observable signatures at Lepton Number and Flavour
Violation experiments. The importance of the fully flavour-covariant
rate equations, as developed in \cite{Dev:2014laa}, for describing accurately the
generation of the lepton asymmetry is reconfirmed.
\end{abstract}
\end{frontmatter}

In this note, we discuss the viability of minimal radiative Resonant
Leptogenesis (RL) scenarios. In these models, the mass splitting between the 
right-handed (RH) heavy Majorana neutrinos, which can explain the observed
baryon asymmetry of the Universe, as well as the low-energy neutrino data, is
generated \emph{entirely} by the renormalization-group (RG) running
from some high mass scale $\mu_X$ (of the order of the grand unification scale) down
to the relevant heavy-neutrino mass scale $m_N$. We take the latter to be of
the order of the electroweak scale, so that it can be tested in current and future experiments. This is the scenario used in the
numerical examples of~\cite{Dev:2014laa}. Here we amend some
incorrect discussions presented in Sections~5 and~6 of this article.
The incorrectness of some of the numerical results given there is
related to the usage, in the numerical analysis, of the incorrect
formulae (2.9) and (2.13) of~\cite{Deppisch:2010fr},  reported in
(5.10) of~\cite{Dev:2014laa}.

\section{No-go theorem for minimal radiative RL at $\mathcal{O}(h^4)$}\label{sec:nogo}

 The relevant heavy-neutrino Lagrangian is given by
%
\begin{eqnarray}
  -\mathcal{L}_N  \  = \  h_{l}{\alpha}  \overline L^{l}
  \widetilde{\Phi}  N_{\mathrm{R}, \alpha}
  + \frac{1}{2} \overline{N}_{\mathrm{R}, \alpha}^C  [M_N]^{\alpha \beta}
  N_{\mathrm{R}, \beta} + \mbox{H.c.}\;,
  \label{eq:L}
\end{eqnarray}
where $\widetilde{\Phi}=i\sigma_2\Phi^*$ is the isospin conjugate of
the Higgs doublet $\Phi$ and the superscript $C$ denotes charge
conjugation. In minimal radiative scenarios, the masses of  these heavy
neutrinos $N_\alpha$  ($\alpha=1,2,3$) are assumed to be degenerate at
a high scale $\mu_X \sim 10^{16}~\mbox{GeV}$, thanks to an
approximate $O(3)$ symmetry: $\mat{M}_N(\mu_X) \ = \ m_N\,\mat{1}_{3}
$. At the scale $m_N$, relevant for leptogenesis, the mass matrix
$\mat{M}_N$ is obtained by the RG evolution from $\mu_X$ to $m_N$:
%
\begin{eqnarray}
\label{eq:mass}
  \mat {M}_N \ = \ m_N\mat{1}_{3} \; + \; \mat{\Delta M}^{\mathrm{RG}}_N\; ,
\end{eqnarray}
where, in the minimal radiative RL scenario, $\mat  {\Delta  M}^{\mathrm{RG}}_N$  is taken to be the only  $O(3)$-breaking correction to the mass
matrix and is given by
%
\begin{eqnarray}
 \mat{\Delta  M}_N^{\mathrm{RG}} \ \simeq \ - \,\frac{m_N}{8\pi^2}
  \ln\left(\frac{\mu_X}{m_N}\right)
  \operatorname{Re}\left[\mat{h}^\dag(\mu_X) \mat{h}(\mu_X)\right] \; .
  \label{deltam}
\end{eqnarray}

However, as we are going to show below, this minimal scenario is not
viable at $\mathcal{O}(h^4)$, because of the following \emph{no-go
theorem for minimal  radiative \emph{RL} at $\mathcal{O}(h^4)$}. The
RH neutrino mass matrix given by~\eqref{eq:mass}
and~\eqref{deltam} is real and symmetric and, as long as we are in the
perturbative regime $|[\Delta M_N^{\mathrm{RG}}]_{\alpha \beta}|/m_N \ll 1$,
it can be diagonalized with positive eigenvalues by a real orthogonal
matrix $\mat{O} \in O(3) \subset U(3)$:
%
\begin{eqnarray}
\mat{M}_N \ = \ \mat{O} \, \widehat{\mat{M}}_N \, \mat{O}^{\mathsf{T}}
\;,
\end{eqnarray}
where the caret ( $ \widehat{} $ ) denotes the mass eigenbasis. At
leading order, i.e.~$\mathcal{O}(h^2)$, the Yukawa couplings
in~\eqref{deltam} can be taken at the scale $m_N$. Since $\mat{O}$ is
real and orthogonal, both $\mat{O}^{\mathsf T} \mat{\Delta M}_N^{\mathrm{RG}} \mat{O}$ and
%
\begin{eqnarray}
{\rm Re}(\widehat{\mat{h}}^\dagger \widehat{\mat{h}}) \ = \ {\rm Re}\big[
(\mat{O}^{\mathsf{T}} \mat{h}^\dagger)(\mat{h} \mat{O})\big] \ =
\ \mat{O}^{\mathsf{T}} \, {\rm Re}(\mat{h}^\dagger \mat{h}) \, \mat{O} \
\propto \ \mat{O}^{\mathsf T} \mat{\Delta M}_N^{\mathrm{RG}} \mat{O}
\end{eqnarray}
are also separately diagonal. On the other hand, the leptonic asymmetry
$\varepsilon_{l\alpha}$ in the decay $N_\alpha \to L_l \Phi$ is
proportional to the quantity (cf.~(A.2) in~\cite{Dev:2014laa})
%
\begin{eqnarray}
{\rm Im}\big[\widehat{h}^*_{l\alpha}
\widehat{h}_{l\beta}(\widehat{h}^\dag
\widehat{h})_{\alpha\beta}\big]+\frac{m_{N,\,\alpha}}{m_{N,\,\beta}}
\operatorname{Im}\big[\widehat{h}^*_{l\alpha}
\widehat{h}_{l\beta}(\widehat{h}^\dag
\widehat{h})_{\beta\alpha}\big] 
\ = \ 2
\operatorname{Im}\big[\widehat{h}^*_{l\alpha}
\widehat{h}_{l\beta}\big] \, {\rm Re}\big[(\widehat{h}^\dag
\widehat{h})_{\alpha\beta}\big] \; + \; \mathcal{O}(h^6)\;,
\end{eqnarray}
where $m_{N,\alpha}$ is the physical mass of $N_\alpha$ and $\alpha
\neq \beta$. \emph{Therefore, the leptonic asymmetry
$\varepsilon_{l\alpha} \propto {\rm Re}\big[(\widehat{h}^\dag
\widehat{h})_{\alpha\beta}\big]$, being proportional to the
off-diagonal entries of a diagonal matrix, vanishes identically at
$\mathcal{O}(h^4)$ in minimal radiative models, where no other source
of $O(3)$ flavour breaking is present.}\qed

\section{A next-to-minimal radiative RL model}\label{sec2}

To avoid the no-go theorem of suppressed leptonic asymmetries, as
derived in Section~\ref{sec:nogo}, we proceed differently
from~\cite{Dev:2014laa,Deppisch:2010fr}. We include a new source of
flavour breaking $\mat{\Delta M}_N$, which is {\em not} aligned with
${\rm Re}(\mat{h}^\dagger \mat{h})$ at the input scale $\mu_X$. More
explicitly, the heavy-neutrino mass matrix takes on the following form:
%
\begin{eqnarray}
  \mat {M}_N \ = \ m_N\mat{1} \; + \; \mat{\Delta M}_N \; + \;
  \mat{\Delta M}^{\mathrm{RG}}_N\; .
\end{eqnarray}
For the purposes of this note, we consider a minimal breaking matrix $\mat{\Delta M}_N$ of the form
%
\begin{eqnarray}
\mat{\Delta M}_N \ = \ \begin{pmatrix}
\Delta M_1 & 0 & 0 \cr
0 & \Delta M_2/2 & 0 \cr
0 & 0 & -\Delta M_2/2
\end{pmatrix} \;,
\end{eqnarray}
where $\Delta M_2$ is needed to make the light-neutrino mass matrix
rank-2, thus allowing us to fit successfully the low-energy neutrino
data. On the other hand, $\Delta M_1$ governs the mass difference between $N_1$
and $N_{2,3}$, and its inclusion is sufficient to obtain successful
leptogenesis.

In order to protect the lightness of the left-handed neutrinos in  a
technically natural manner, we consider an $\mathrm{RL}_\tau$ model that
possesses a leptonic symmetry $U(1)_l$. In this scenario, the Yukawa
couplings $h_{l}{\alpha}$ have the following
structure~\cite{Pilaftsis:2004xx,Pilaftsis:2005rv}:
%
\begin{eqnarray}
  \mat{h} \ = \begin{pmatrix}      0 & a \,e^{-i\pi/4} & a\,e^{i\pi/4}\cr
      0 & b\,e^{-i\pi/4} & b\,e^{i\pi/4}\cr
      0 & c\,e^{-i\pi/4} & c\,e^{i\pi/4}\end{pmatrix}
     \: + \: \mat{\delta h} \; ,
  \label{yuk}
\end{eqnarray}
where, in order to protect the $\tau$ asymmetry from excessive
washout and at the same time guarantee observable effects in low-energy
neutrino experiments, we take $|c| \ll |a|,|b| \approx
10^{-3}-10^{-2}$. The leptonic flavour-symmetry-breaking matrix is
taken to be
%
\begin{eqnarray}
  \mat{\delta h} \ = \ \begin{pmatrix}
      \epsilon_e & 0 & 0\cr
      \epsilon_\mu & 0 & 0\cr
      \epsilon_\tau & 0 & 0 \end{pmatrix}
 \; .
\label{delta_h}
\end{eqnarray}
To leading order in the symmetry-breaking parameters of $\Delta \mat{
M}_N$ and  $\mat{\delta h}$,  the tree-level  light-neutrino mass
matrix is given by the seesaw formula
%
\begin{eqnarray}
  \bm{M}_\nu \ \simeq \ -\frac{v^2}{2}\mat{h} \bm{M}_N^{-1} \mat{h}^{\mathsf{T}} \
  \simeq \ \frac{v^2}{2m_N}\begin{pmatrix}
      \frac{\Delta m_N}{m_N} a^2 - \epsilon_e^2 &
     \frac{\Delta m_N}{m_N} ab - \epsilon_e\epsilon_\mu &
      - \epsilon_e\epsilon_\tau\cr
      \frac{\Delta m_N}{m_N} ab - \epsilon_e\epsilon_\mu &
      \frac{\Delta m_N}{m_N} b^2 - \epsilon_\mu^2 &
      - \epsilon_\mu\epsilon_\tau\cr
      - \epsilon_e\epsilon_\tau &
      - \epsilon_\mu\epsilon_\tau &
      - \epsilon_\tau^2 \end{pmatrix}
  ,
  \label{mnu}
\end{eqnarray}
where
%
\begin{eqnarray}
\Delta m_N \equiv 2 \,[\Delta M_N]_{23} + i\,\big([\Delta
M_N]_{33} - [\Delta M_N]_{22}\big) \ = \ -\, i\, \Delta M_2
\end{eqnarray}
and we have neglected subdominant terms $\frac{\Delta m_N}{m_N} \,c
\times (a,b,c)$. Assuming a particular mass hierarchy between the
light-neutrino masses $m_{\nu_i}$   and   for   given   values  of
the   $\mathit{CP}$ phases $\delta,\varphi_{1,2}$,  we determine the following
model parameters appearing in the Yukawa coupling matrix~\eqref{yuk}:
%
\begin{align}
  a^2 \ &= \ \frac{2m_N}{v^2}
  \left(M_{\nu,{11}}-\frac{M^2_{\nu,{13}}}{M_{\nu,{33}}}\right) \frac{m_N}{\Delta m_N}\; , 
\qquad   
  b^2 \ = \ \frac{2m_N}{v^2}
  \left(M_{\nu,{22}}-\frac{M^2_{\nu,{23}}}{M_{\nu,{33}}}\right)\frac{m_N}{\Delta m_N}\; ,
  \nonumber \\[6pt]
  \epsilon_e^2 \  &= \ -\frac{2m_N}{v^2}\frac{M^2_{\nu,{13}}}{M_{\nu,{33}}} \; ,
\qquad  
  \epsilon_\mu^2 \ = \ -\frac{2m_N}{v^2}\frac{M^2_{\nu,{23}}}{M_{\nu,{33}}}\; ,
\qquad  
  \epsilon_\tau^2 \ = \ -\frac{2m_N}{v^2}M_{\nu,{33}}\; . 
  \label{esq} 
\end{align}
Therefore, the Yukawa  coupling matrix~\eqref{yuk} in  the
RL$_\tau$ 
model can be  completely fixed in terms of the  heavy-neutrino mass
scale $m_N$   and  the   input  parameters   $c$ and $\Delta M_{2}$.
Notice that, whereas~\eqref{mnu} and~\eqref{esq} coincide formally with
the corresponding formulae in~\cite{Deppisch:2010fr,Dev:2014laa}, the
latter are incorrect for the model considered therein.

We amend the three benchmark points considered in~\cite{Dev:2014laa} as
detailed in Table~\ref{tab3}. The input parameters $\Delta M_1$ and $c$
are easily chosen such that leptogenesis is successful. But
$\Delta M_2$ has been tuned here in order to reproduce \emph{exactly} the
predictions for the Lepton Number and Flavour Violation (LNV and LFV)
observables discussed in~\cite{Dev:2014laa}. In particular, Table 4
of~\cite{Dev:2014laa} is unaltered, thus confirming the observable
effects in LNV and LFV experiments predicted by this class of models,
while simultaneously allowing for successful leptogenesis. The $\mathit{CP}$
phases in the light neutrino sector have been chosen as $\varphi_1 = - \pi$ and
$\varphi_2=\delta=0$.

\begin{table}[t!]
  \begin{center}
    \begin{tabular}{c|c|c|c}\hline\hline
      Parameters & BP1 & BP2 & BP3\\ \hline\hline
      $m_N$ & 120 GeV & 400 GeV & 5 TeV \\
      $c$ & $2 \times 10^{-6}$ & $2 \times 10^{-7}$ & $2 \times 10^{-6}$ \\
      $\Delta M_1/m_N$ & $-\, 5 \times 10^{-6}$ & $-\, 3 \times 10^{-5}$ & $-\, 4 \times 10^{-5}$ \\ 
      $\Delta M_2/m_N$ & $(- 1.59-0.47\,i)  \times 10^{-8}$ & $(- 1.21+0.10\,i)  \times 10^{-9}$ & $(- 1.46+0.11\,i)  \times 10^{-8}$ \\
       \hline
      $a$ & $(5.54 -  7.41\, i)\times 10^{-4}$ &
      $(4.93-2.32 \, i)\times 10^{-3}$ &
      $(4.67-4.33 \, i)\times 10^{-3}$\\
      $b$ & $(0.89 - 1.19\, i)\times 10^{-3}$ &
      $(8.04 - 3.79 \, i)\times 10^{-3}$ &
      $(7.53-6.97 \, i)\times 10^{-3}$\\
      $\epsilon_e$ & $3.31\,i\times 10^{-8}$ &
      $5.73\, i\times 10^{-8}$ & $2.14\, i\times 10^{-7}$ \\
      $\epsilon_\mu$ & $2.33\, i\times 10^{-7}$ &
      $4.30\, i\times 10^{-7}$ & $1.50\, i\times 10^{-6}$ \\
      $\epsilon_\tau$ & $3.50\, i\times 10^{-7}$ &
      $6.39\, i\times 10^{-7}$ & $2.26\, i\times 10^{-6}$ \\  
      \hline\hline
    \end{tabular}
  \end{center}
  \caption{The   numerical    values   of   the   free   ($m_N$,
$c$,   $\Delta M_{1,2}$)  and   derived
parameters ($a$,  $b$, $\epsilon_{e,\mu,\tau}$)  
in  the RL$_\tau$ model
for three chosen benchmark points.}
  \label{tab3}
\end{table}

The discussion in Section~5.3.2 of~\cite{Dev:2014laa}, concerning the
approximate analytic solution for the charged-lepton decoherence
effect, also requires modifications. In particular, some of the
approximations adopted there are no longer valid. In light of this,
(5.22) of~\cite{Dev:2014laa} becomes
%
\begin{eqnarray}
\label{eq:eta_L_evol_simpl}
  \frac{\D{}{}}{\D{}{z}} [\delta \widehat{\eta}^L]_{lm} \ & = \
  \frac{z^3 K_1(z)}{2} \bigg(\sum_\alpha \: [\widehat{\upeta}^N]_{\alpha \alpha} \,
  [\delta \widehat{\mathrm{K}}^N_{L \Phi}]_{l m \alpha \alpha} \
  - \ \frac{1}{3} \, \Big\{\delta \widehat{\eta}^{L} , \,
  \widehat{\mathrm{K}}^{\mathrm{eff}}\Big\}_{lm} \nonumber\\
  &- \ \frac{2}{3} \, [\delta \widehat{\eta}^{L}]_{kn} \, [\widehat{\mathrm{K}}_{-}]_{nklm}\; - \; \frac{2}{3} \,
  \Big\{\delta \widehat{\eta}^L, \,
  \widehat{\mathrm{K}}_{\mathrm{dec} } \Big\}_{l m} \;
  +\; [\delta \widehat{\mathrm{K}}_{\mathrm{dec}}^{\rm{back}}]_{ l m}  \bigg) \;,
\end{eqnarray}
where $\{,\}$ denotes anti-commutators in flavour space and we need to
  introduce also the K-factor $[\widehat{\mathrm{K}}_{-}]_{nklm} = \upkappa
  \big[ \widehat{{\gamma}}^{L\Phi}_{L^{\tilde{c}}
  \Phi^{\tilde{c}}} - \widehat{{\gamma}}^{L\Phi}_{L \Phi}
  \big]_{nklm}$, which is no longer subdominant.
  Correspondingly, (5.26) of~\cite{Dev:2014laa} is modified to
%
\begin{eqnarray}
  \frac{1}{3} \, \Big\{ {\delta} \widehat{{\eta}}^{L} , \,
  \widehat{\mathrm{K}}^{\mathrm{eff}} + 2 \,\widehat{\mathrm{K}}_{\mathrm{dec}}
  \Big\}_{lm} \;
  -\; [{\delta} \widehat{\mathrm{K}}_{\mathrm{dec}}^{\rm{back}}]_{lm} \;+
  \;\frac{2}{3} \, [\delta \widehat{\eta}^{L}]_{kn} \, [\widehat{\mathrm{K}}_{-}]_{nklm} \
  \simeq \ \frac{\widehat{{\varepsilon}}_{lm}}{z} \; .
  \label{attractor_lm}
\end{eqnarray}
It is not easy to perform further approximations in this equation, and
therefore, it is convenient to solve numerically the linear system~\eqref{attractor_lm} for the variables
$[\delta \widehat{\eta}^L]_{lm}$. We then obtain the
semi-analytic contribution of mixing and charged-lepton (de)coherence
to the asymmetry in the strong-washout regime
%
\begin{eqnarray}
\label{eq:analytic} \delta \widehat{\eta}^L \ \supset \ \delta
\widehat{\eta}^L_{\mathrm{mix}} + \delta \widehat{\eta}^L_{\mathrm{dec}} \ \simeq
\ \sum_l \, [\delta \widehat{\eta}^L]_{ll} \;,
\end{eqnarray}
where the diagonal asymmetries $[\delta \widehat{\eta}^L]_{ll}$ are
obtained by solving the linear system~\eqref{attractor_lm}.

\begin{figure}[p!]
  \centering
\vspace{-1.5cm}
 \includegraphics[width=10cm]{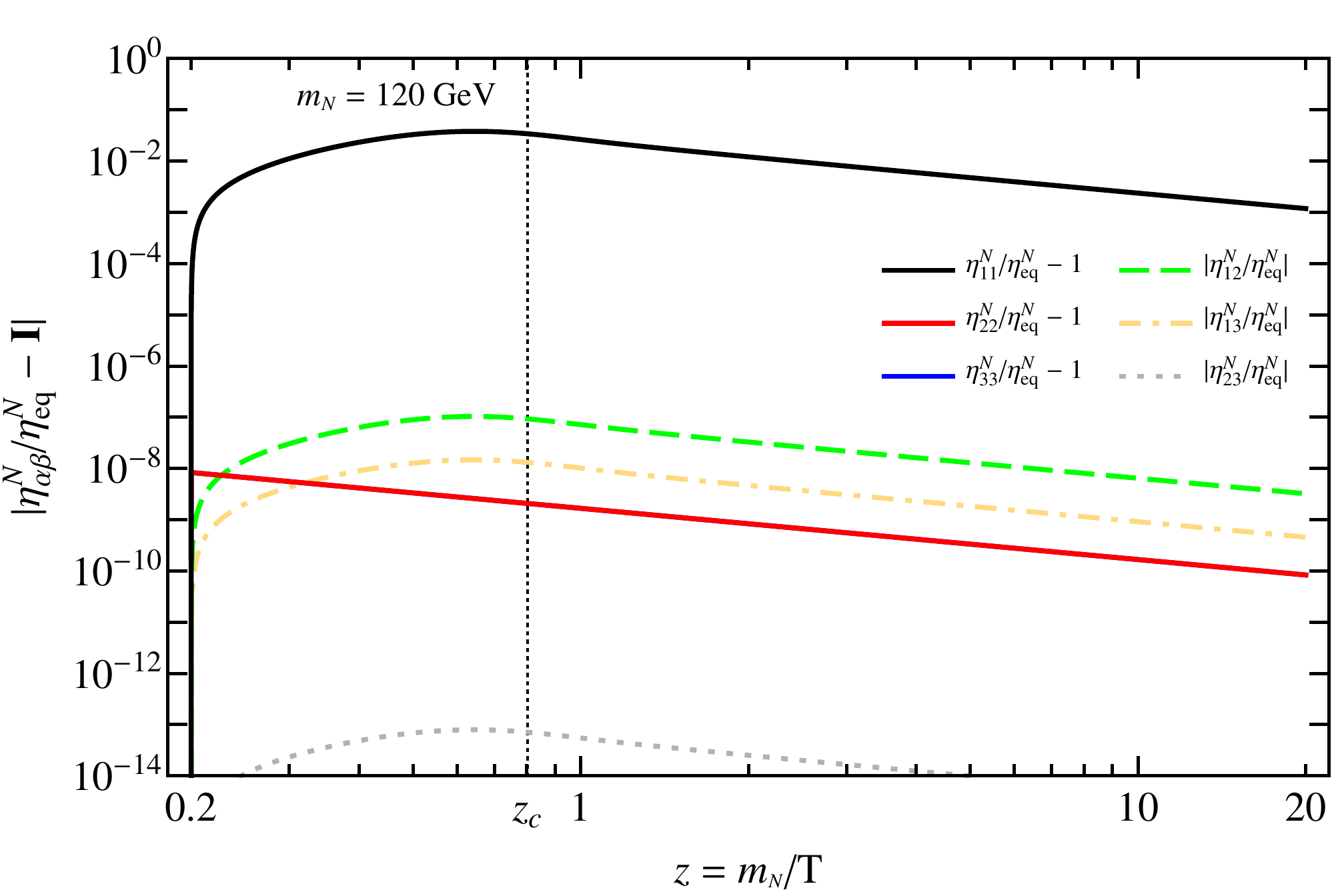}
\\
 \includegraphics[width=10cm]{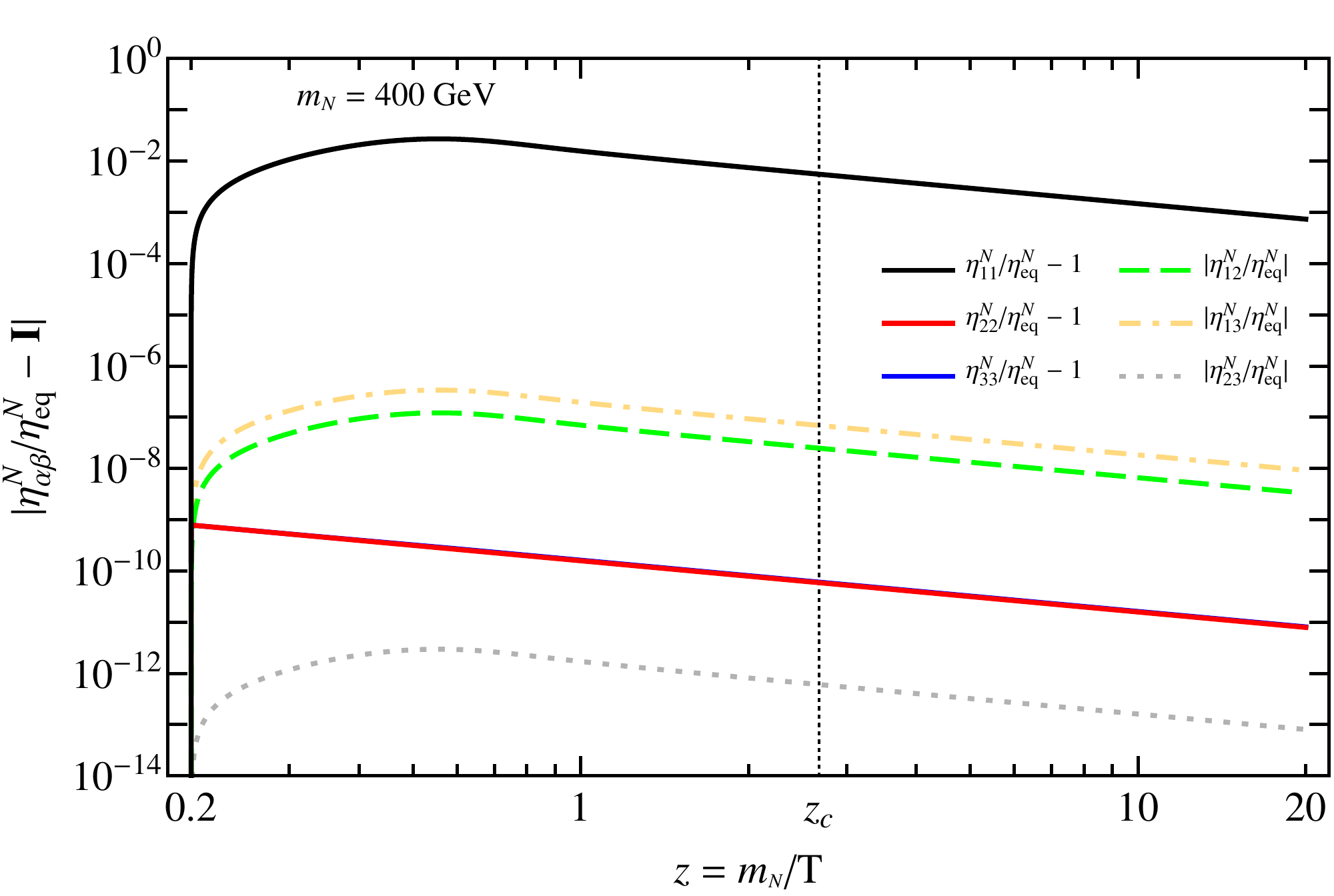} \\
\includegraphics[width=10cm]{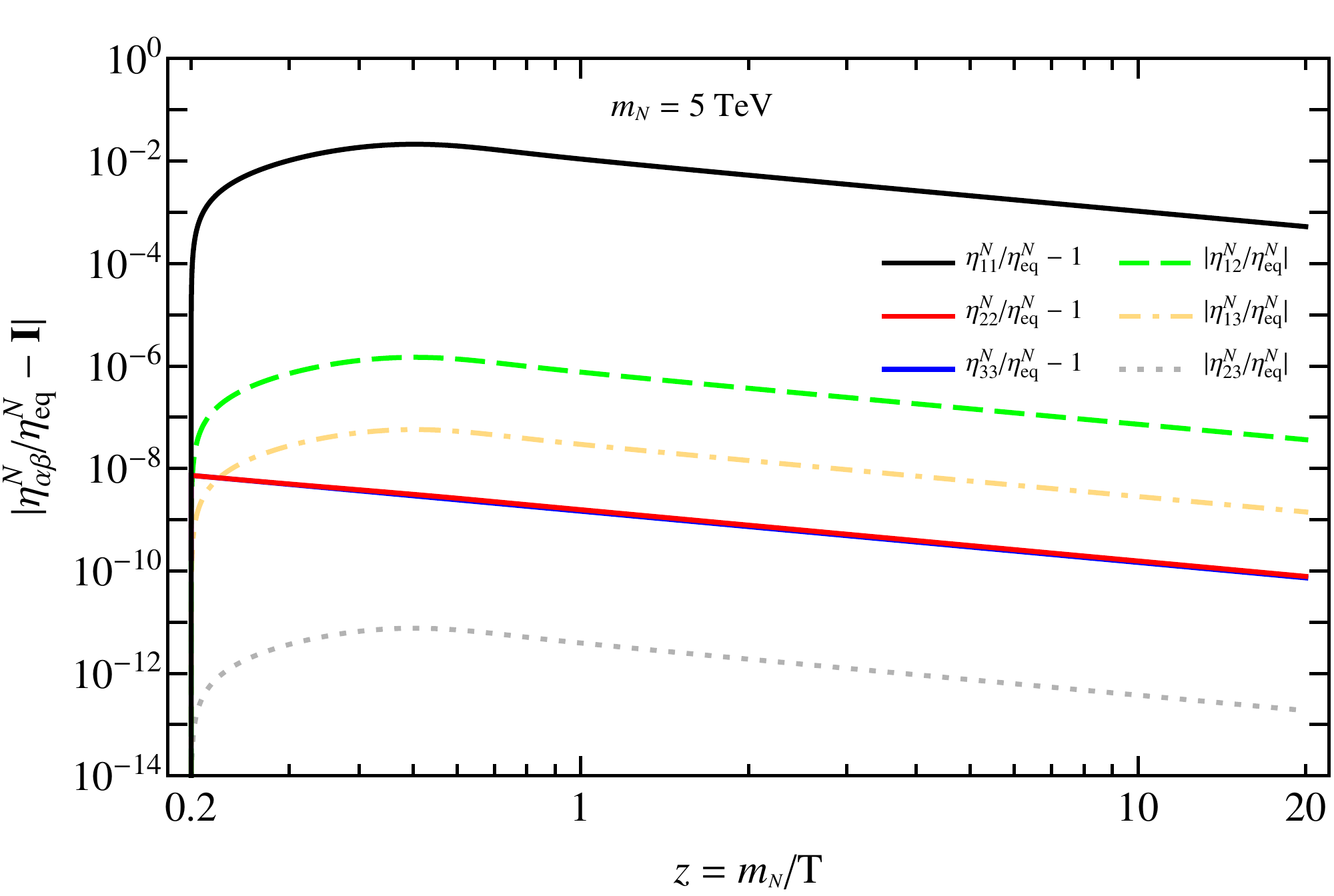} 
  \caption{The deviation of the heavy-neutrino number densities
$\widehat{\upeta}^N_{\alpha\beta}=\widehat{\eta}^N_{\alpha\beta}/\eta^N_{\mathrm{eq}}-\delta_{\alpha\beta}$
from their equilibrium values for  the three benchmark points given in
Table~\ref{tab3}.   The  different lines  show  the  evolution of  the
diagonal  (solid   lines)  and  off-diagonal   (dashed  lines)  number
densities  in  the fully  flavour-covariant  formalism.  The  numerical
values of $\widehat{\upeta}^N_{22}$ and $\widehat{\upeta}^N_{33}$ coincide with each other
in all three cases.  }
\label{figN}
\end{figure}

\begin{figure}[p!]
  \centering
\vspace{-1.5cm}
 \includegraphics[width=14.cm]{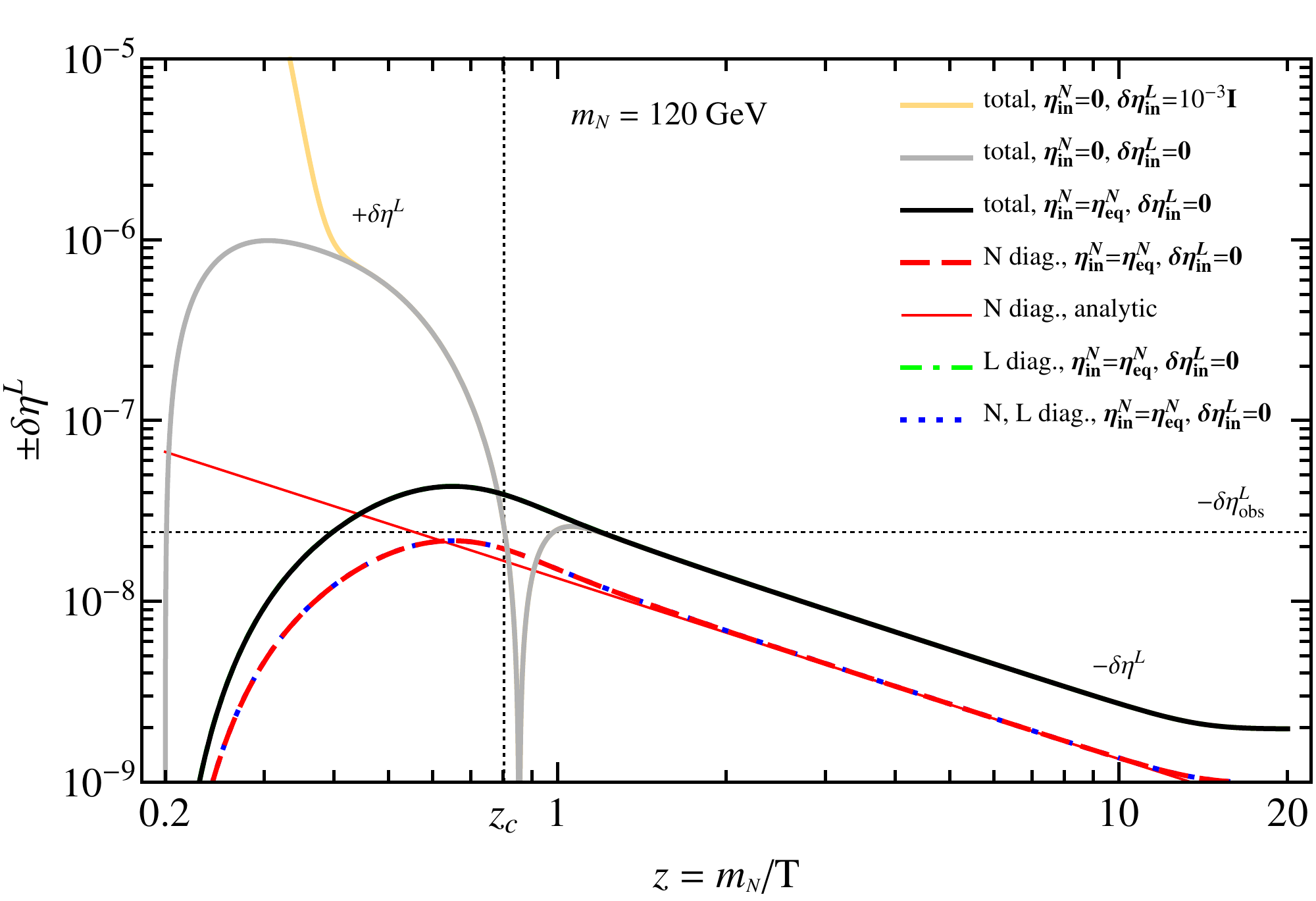}
\\
 \includegraphics[width=14.cm]{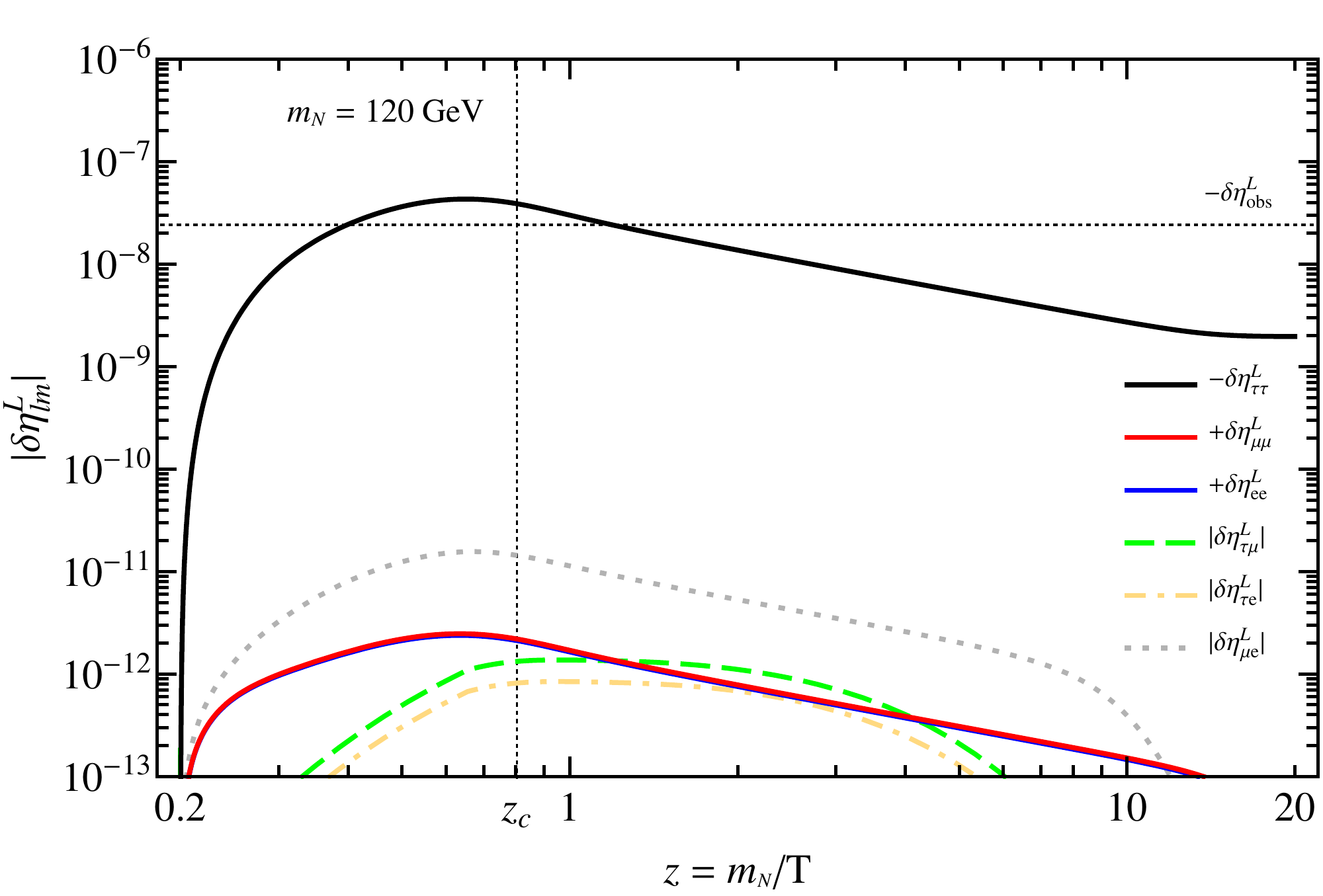}
  \caption{Lepton  flavour asymmetries as predicted  by the {\rm BP1}
parameters  given  in  Table~\ref{tab3}.   The  top  panel  shows  the
comparison  between  the  total  asymmetry obtained  using  the  fully
flavour-covariant formalism (thick solid lines, with different initial
conditions) with  those obtained using  the flavour-diagonal formalism
(dashed lines).  Also shown (thin  solid line) is the  semi-analytic result~\eqref{eq:analytic}.   The  bottom  panel  shows  the
diagonal (solid lines) and off-diagonal (dashed lines) elements of the
total lepton  number asymmetry  matrix in the  fully flavour-covariant
formalism. $\delta \widehat{\eta}^L_{ee}$ and $\delta \widehat{\eta}^L_{\mu \mu}$ are coincident. For details, see the text and~\cite{Dev:2014laa}.}
\label{fig5}
\end{figure}

\begin{figure}[p!]
  \centering
  \includegraphics[width=14.cm]{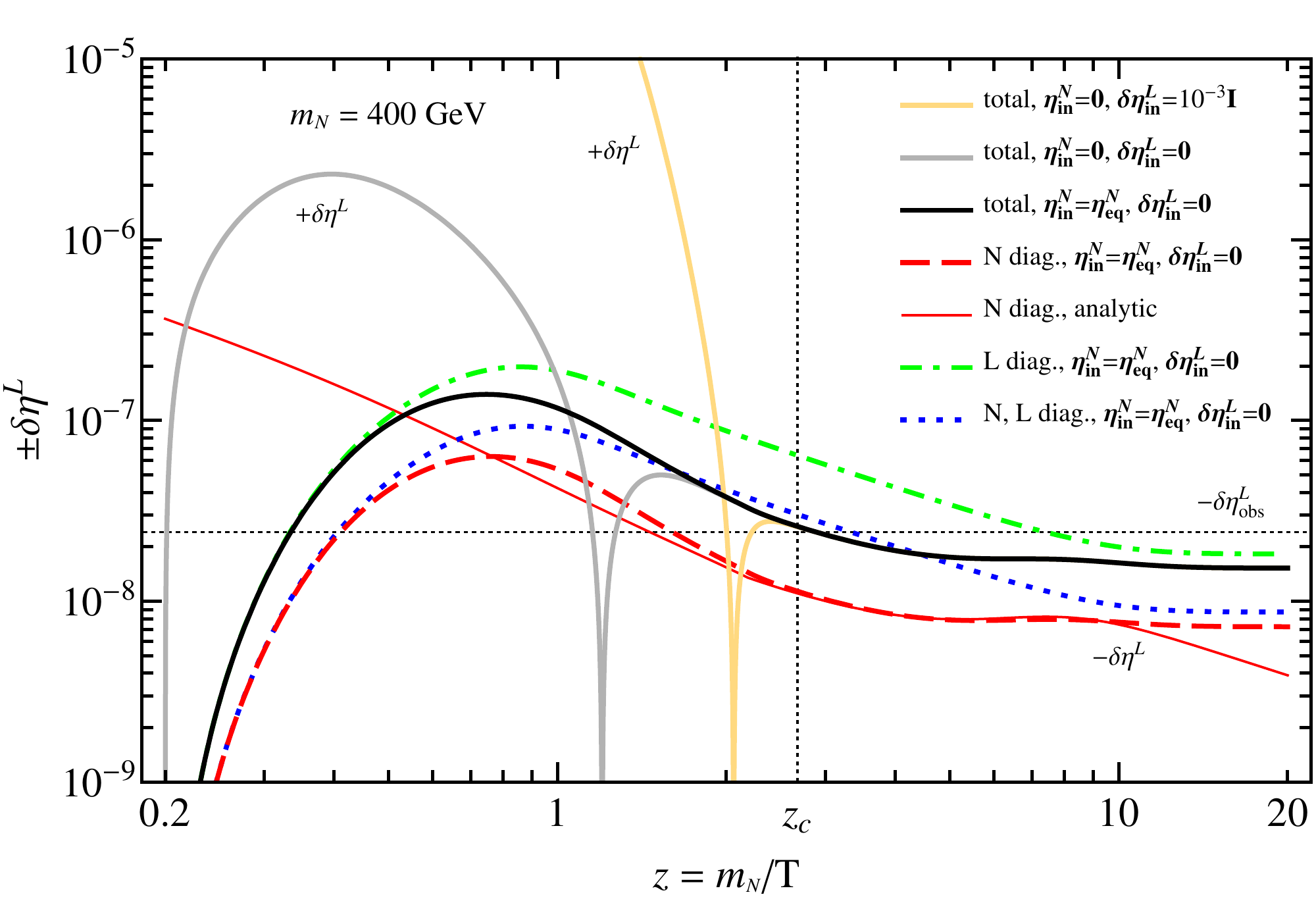}\\
  \includegraphics[width=14.cm]{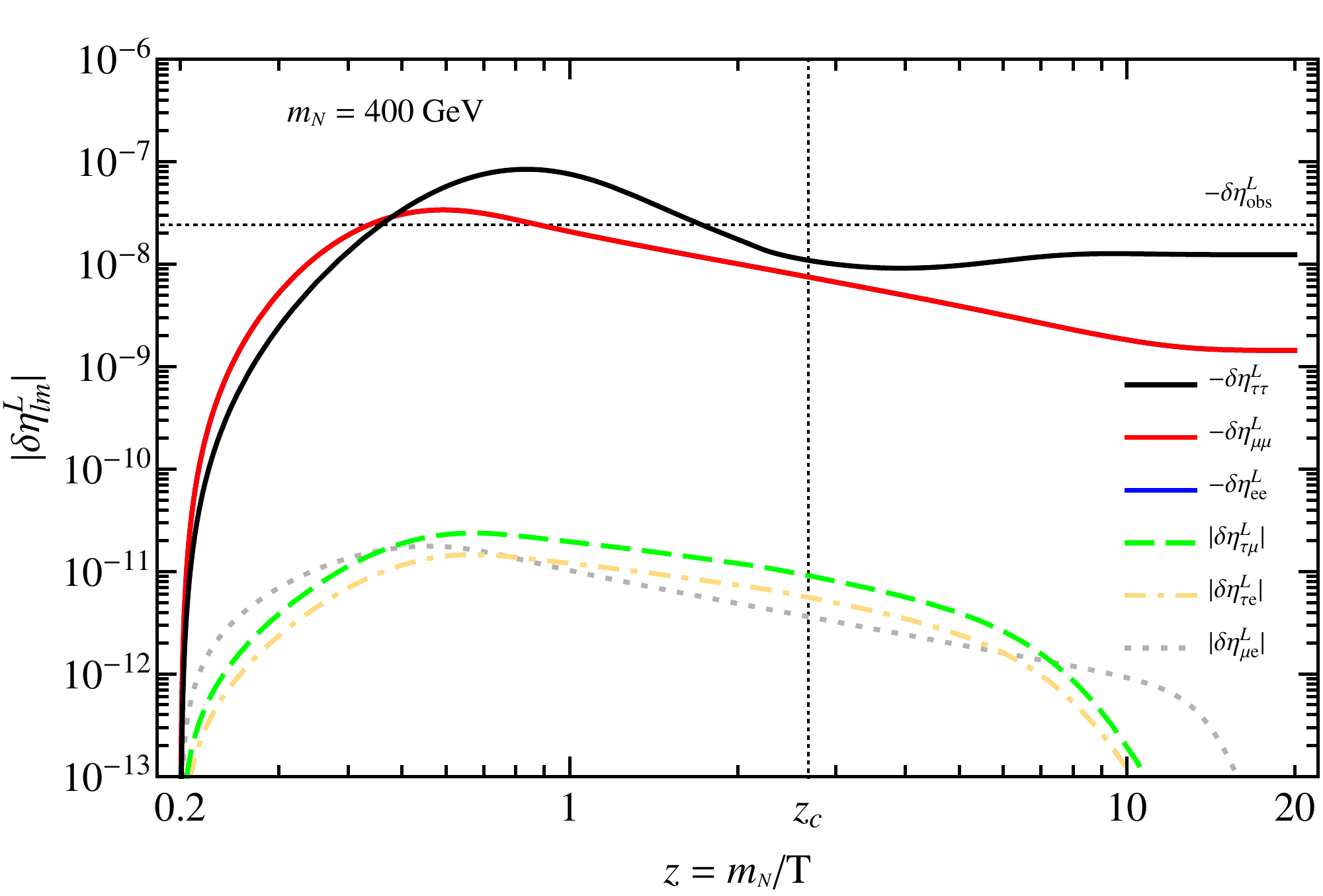}
    \caption{Lepton flavour asymmetries as  predicted by
the {\rm BP2 RL$_\tau$}  model parameters given in Table~\ref{tab3}.
The labels are the same as in Figure~\ref{fig5}. $\delta \widehat{\eta}^L_{ee}$ and $\delta \widehat{\eta}^L_{\mu \mu}$ are coincident. \vspace{2em}}
  \label{fig6}
\end{figure}

\begin{figure}[p!]
  \centering
  \includegraphics[width=14.cm]{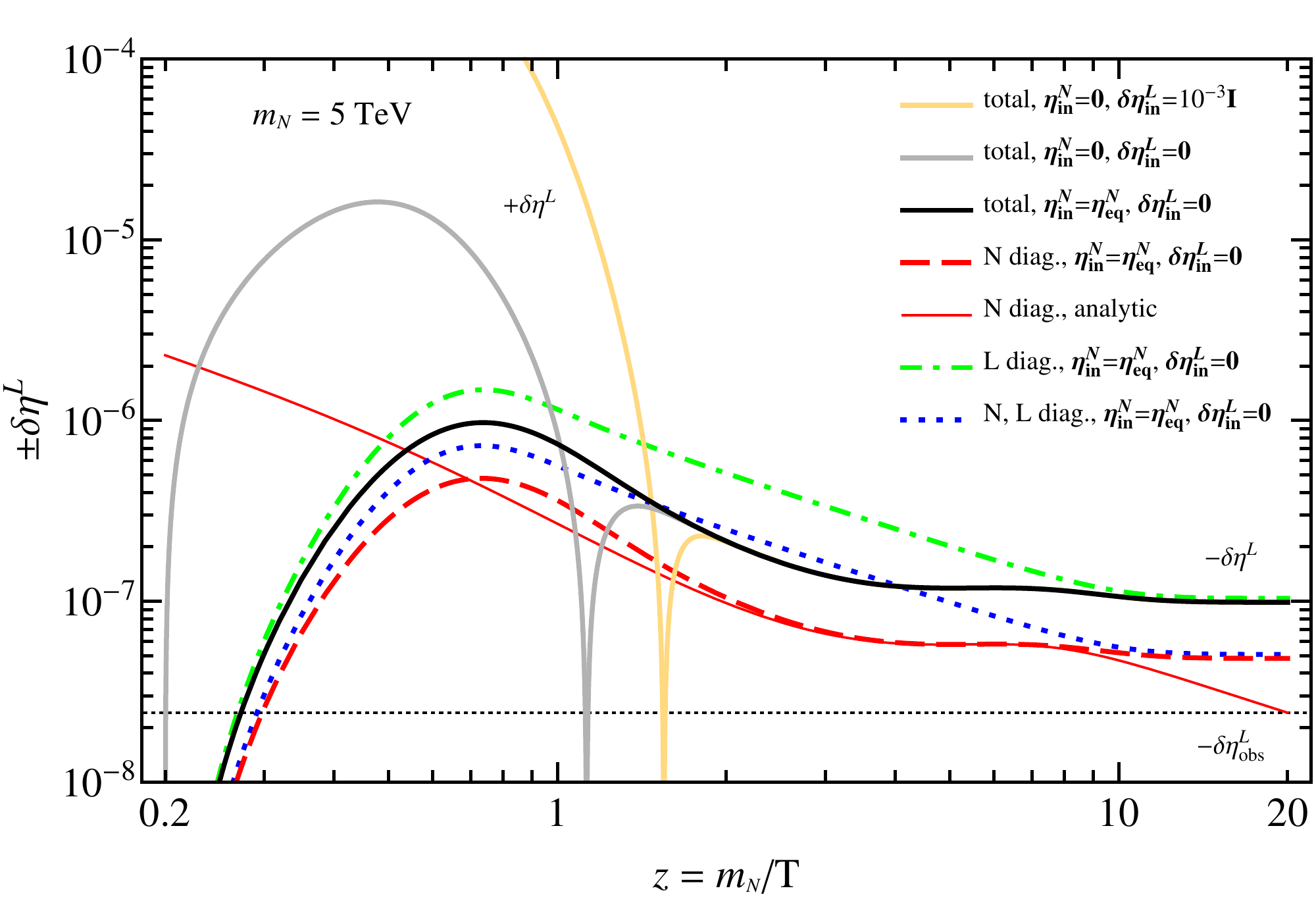}\\
  \includegraphics[width=14.cm]{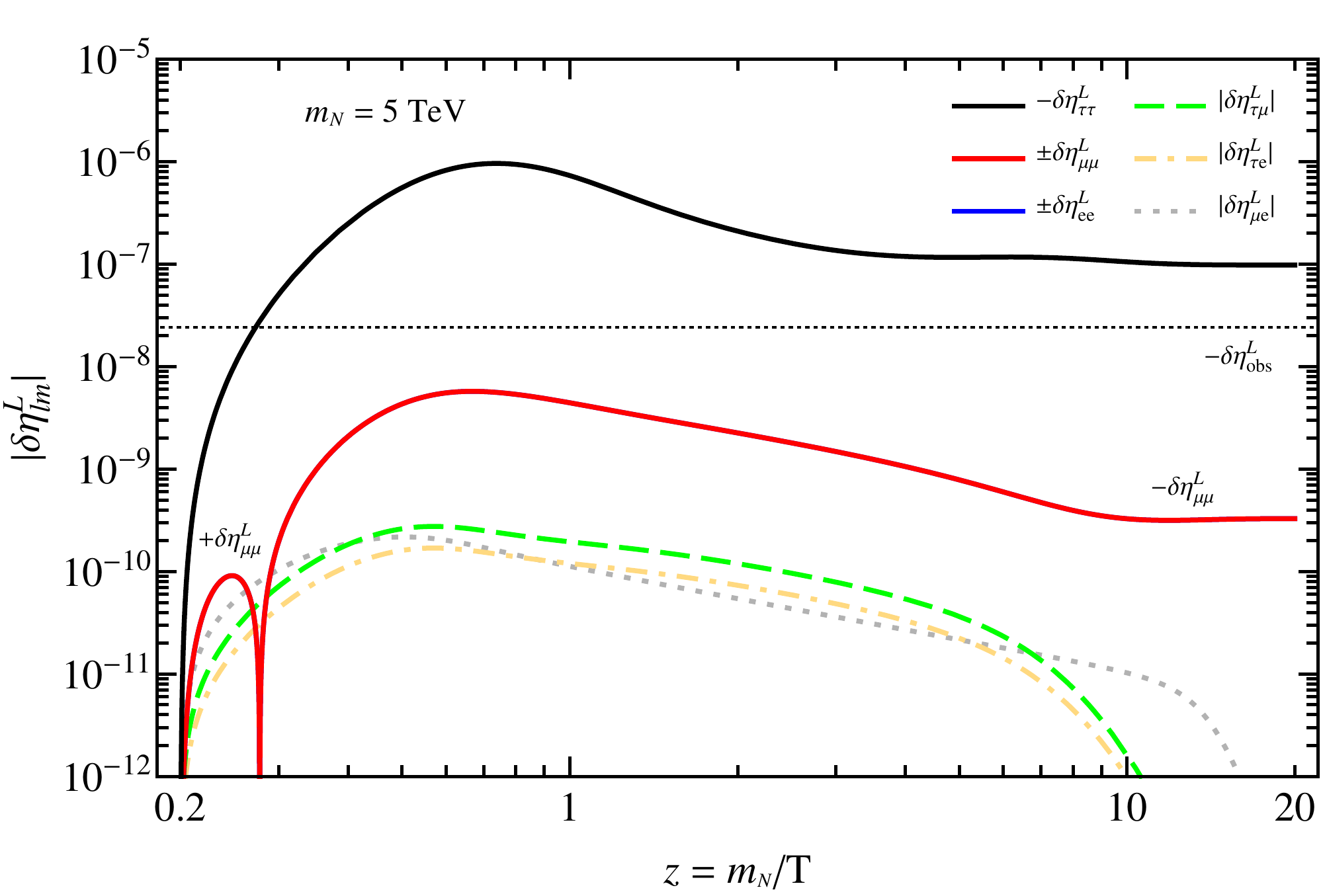}
  \caption{Lepton flavour asymmetries as predicted by the
{\rm BP3 RL$_\tau$} model parameters given in Table~\ref{tab3}.  The
labels are the same as in Figure~\ref{fig5}. $\delta \widehat{\eta}^L_{ee}$ and $\delta \widehat{\eta}^L_{\mu \mu}$ are coincident. \vspace{2em}}
  \label{fig7}
\end{figure}

Finally, Figs.~8--11 of~\cite{Dev:2014laa} are modified too. The
amended numerical results are shown in
Figs.~{figN-fig7} of this note. The
main qualitative difference with respect to those given
in Section 6.2 of~\cite{Dev:2014laa} is that the contribution of the charged-lepton
off-diagonal number densities now \emph{suppresses} the total asymmetry
for the three benchmark points considered here, rather than enhancing
it, as in Figs.~10--11 of~\cite{Dev:2014laa}. Nevertheless, successful
leptogenesis is still comfortably realized. Thus, we may conclude that
the salient features discussed in Sections 5 and 6
of~\cite{Dev:2014laa} remain valid, namely the joint possibility of
successful leptogenesis and observable signatures in LNV and LFV
experiments. Moreover, as is evidenced by the disparity between the
asymmetries predicted by the partially flavour-off-diagonal treatments
in Figs.~\ref{fig6} and \ref{fig7}, the use of fully flavour-covariant
rate equations, as developed in~\cite{Dev:2014laa}, remains of
paramount importance for obtaining accurate quantitative predictions in
this class of models.

\section*{Acknowledgements}
We thank Jonathan Da Silva for pointing out the incorrectness of (2.13) in~\cite{Deppisch:2010fr}, reported as (5.10) in~\cite{Dev:2014laa}.
The work  of P.S.B.D. and  A.P.  is supported  by the
Lancaster-Manchester-Sheffield  Consortium   for  Fundamental  Physics
under  STFC   grant  ST/L000520/1. The work of P.M. is supported by a University Foundation Fellowship (TUFF) from the Technische Universit\"{a}t M\"{u}nchen and the Deutsche Forschungsgemeinschaft (DFG) cluster of
excellence Origin and Structure of the Universe. The work of D.T. has been supported by a fellowship of the EPS  Faculty of  the  University  of Manchester.

%

\end{document}